\documentclass[runningheads,a4paper]{llncs}

\usepackage{amssymb}
\usepackage{graphicx}

\begin{document}

\title{Using a Dynamic Neural Field Model to Explore a Direct Collicular Inhibition Account of Inhibition of Return}

\titlerunning{IOR as Direct Collicular Inhibition}

\author{Jason Satel\thanks{Correspondence should be addressed to J. Satel (J.Satel@dal.ca).}\inst{1} \and Ross Story\inst{1} \and Matthew D. Hilchey\inst{1}
\and \\Zhiguo Wang\inst{2} \and Raymond M. Klein\inst{1}}

\authorrunning{Satel et al.}

\institute{Department of Psychology and Neuroscience, Faculty of Science, Dalhousie University
\and Center for Cognition and Brain Disorders, Hangzhou Normal University}

\toctitle{Neural Field Model of IOR}
\tocauthor{Satel et al.}
\maketitle

\begin{abstract}

When the interval between a transient flash of light (a ``cue" ) and a second visual response signal (a ``target") exceeds at least 200 ms, responding is slowest in the direction indicated by the first signal. This phenomenon is commonly referred to as inhibition of return (IOR). The dynamic neural field model (DNF) has proven to have broad explanatory power for IOR, effectively capturing many empirical results. Previous work has used a short-term depression (STD) implementation of IOR, but this approach fails to explain many behavioral phenomena observed in the literature. Here, we explore a variant model of IOR involving a combination of STD and delayed direct collicular inhibition. We demonstrate that this hybrid model can better reproduce established behavioural results. We use the results of this model to propose several experiments that would yield particularly valuable insight into the nature of the neurophysiological mechanisms underlying IOR.

\end{abstract}

\section{Introduction}

Inhibition of return (IOR) commonly refers to an extended period (about 3 s, e.g., \cite{vaughan84,samuel03}) of slowed responding toward and/or at the location of a spatially irrelevant visual signal soon after its onset \cite{posner85} (see \cite{klein00}, for review). The phenomenon has been extensively studied in the spatial cueing paradigm \cite{posner80,posner84}. In this paradigm, the interval between two visual onset signals is often manipulated. Conventionally, the first and second signals are referred to as the ``cue" and ``target", respectively, and the interval between them is called the cue-target onset asynchrony (CTOA). IOR is robust at CTOAs greater than 300 ms but can also be detected as early as 50 ms post-cue onset \cite{danziger99}. IOR plays an important role in visual search by biasing responding against previously inspected regions in space \cite{posner85,klein99} and may have evolved to optimize foraging behaviors \cite{klein88}. 

\subsection{Neural Origins of IOR}
The relative contribution of cortical and subcortical oculomotor processing mechanisms to IOR have been the subject of intense debate and controversy. IOR exists primarily in spatiotopic or environmental coordinates when dissociated from retinotopic reference frames \cite{maylor85,posner84}, an important property of IOR if it is to function effectively as a foraging facilitator \cite{hilchey12NSL}. The contribution of cortical processes, in particular the posterior parietal lobes, to spatiotopic IOR are well-established \cite{sapir04,vankoningsbruggen09,mirpour09} and appear necessary given that low-level oculomotor circuitry is retinotopically-organized. Tipper and colleagues \cite{jordan98,tipper91} provided the seminal demonstration that IOR could also exist in object-based coordinates in addition to having demonstrated the contribution of cortical processes to object-based IOR \cite{tipper97}  (cf. \cite {vivas03}, for evidence that the parietal lobe plays a role in object-based IOR). Work from Sumner and colleagues \cite{sumner04} has revealed that IOR can still be generated to stimuli which initially bypass the superior colliculus (SC), but only in manual and not oculomotor response conditions (see also \cite {bourseous12,smith12}, for  non-collicular origins of IOR). 

On the other hand, a variety of studies have provided strong evidence for the central role of the SC in generating IOR \cite{friedrich85,posner85,sapir99,sereno06}. Simply, even in tasks requiring only manual responding, IOR is abolished in patients with SC lesions or degenerative conditions disrupting normal reflexive oculomotor functioning. Moreover, evidence for IOR has been observed in the archerfish, in which cortical processes are markedly underdeveloped \cite{gabay13}. Converging evidence for the role of low-level oculomotor processes comes from single unit recording studies of the primate SC (cf. \cite{fecteau05}). These studies have identified the intermediate layers of the superior colliculus (iSC) as a probable locus for at least some of the underlying mechanisms of IOR \cite{dorris02,fecteau05}. Dorris et al. \cite{dorris02} examined monkeys trained on a simple IOR task by recording from visual and visuomotor neurons residing in the superficial layers of the SC (sSC) and the iSC, respectively, while the task was performed. These researchers found that there was a reduction in target-elicited activativation at the cued location that was correlated with behaviorally measured saccadic reaction times (SRTs). In another experiment, instead of probing an oculomotor response with a visual target, on 25\% of the trials Dorris et al. evoked an oculomotor response by delivering a train of microstimulation directly to visuomotor neurons. At the 200 ms CTOA, electrically stimulated oculomotor responses were faster to cued as compared to uncued locations whereas no statistical effect was observed at the 1100 ms CTOA. These findings suggested that the reduction in activity arises from upstream sensory afferents. However, their monkeys neither exhibited IOR at long CTOAs when oculomotor responses were electrically-evoked (as noted) nor when made to visual stimulation. As such, it is still possible that local inhibition in the SC arises at later times after cue onset. 

\subsection{The STD and DS theory of IOR}
There are a variety of hypothesized models of IOR but here we focus on two computationally explicit theories. The early sensory adaptation or short-term depression (STD) theory and the local inhibition or direct suppression (DS) theory of IOR. STD theory, as presented by Satel, Wang, Trappenberg, and Klein \cite{satel11}, implements IOR as the result of input attenuation of target-elicited early sensory input signals to iSC. There is strong evidence that STD is a component of early IOR in monkeys from single unit recordings of the iSC during traditional cue-target tasks \cite{dorris02,fecteau05}. The previously mentioned reduction in activity in the iSC at cued locations is correlated with SRTs, and can also be observed in the sSC which only receives input from early sensory areas. Thus, the STD theory of IOR predicts that target-elicited visual inputs to the iSC will be reduced in magnitude for some period when presented at a previously cued location, as a function of the time since previous stimulation.

\begin{figure}
\centering
\includegraphics [scale=0.8]{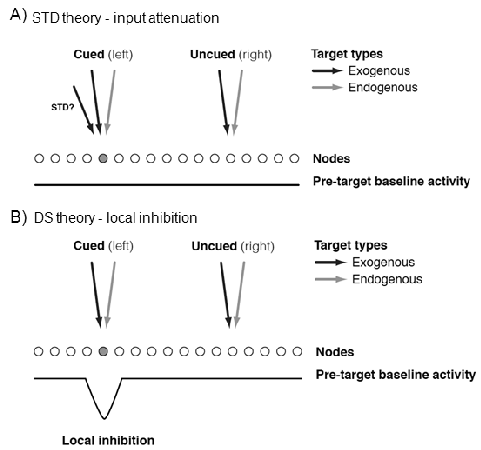}
\caption{Schematic comparison of the A) STD and B) DS theories (adapted from \cite{hilcheyUR}). Exogenous target: peripheral onsets; Endogenous target: arrows at fixation.}
\end{figure}

However, one notable shortcoming of the STD model is its inability to account for IOR when the stimulus commanding an oculomotor response does not occur at a previously stimulated location \cite{taylor00,hilchey12PBR}. Furthermore, it is unclear whether STD could operate in spatiotopic coordinates - a fundamental property of IOR. STD reduces the magnitude of neural responses to stimuli at previously stimulated locations. This reduction in input strength in turn increases the time required for the targeted iSC neuron to reach a firing rate sufficient to initiate a saccade. Yet, as noted, even when an oculomotor response is generated  by left- or right- ward pointing arrows on the fovea, oculomotor responding toward the cued location is delayed. Since STD is not active with arrow stimuli, but IOR is still displayed, some other mechanism must be responsible for the increased buildup time required to generate a saccade to location targeted by a central arrow.

The local inhibition or direct suppression (DS) theory explains IOR in terms of a delayed inhibitory signal centered on a previously stimulated area, reducing the baseline activation level and thus increasing SRTs. As demonstrated by Dorris et al. \cite{dorris02}, such an inhibitory signal is not present in association with IOR in the iSC up to 200 ms post-cue. At longer CTOAs, as aforementioned, the monkeys did not exhibit IOR as measured by oculomotor responses to visual stimulation nor was there any evidence for direct inhibition as revealed by microstimulation. Whereas Dorris et al. \cite{dorris02} failed to observe behavioral evidence for IOR at late CTOAs, behavioral investigations on humans reliably demonstrate IOR at late CTOAs and that its magnitude is similar whether the cue and response signal occur at the same location in space or whether the oculomotor response is commanded by input at fixation\cite{taylor00,hilchey12,hilcheyUR}. Such findings suggest that, at least in humans, a direct inhibitory signal may arrive at the iSC but perhaps later than 200 ms post-cue \cite{hilcheyUR}.

The DS theory alone is unable to explain IOR measured at short CTOAs, but a hybrid STD plus DS theory, predicts behavior well at all CTOAs - with both peripheral onset and central arrow stimuli - forming an effective theoretical framework for understanding many of the behavioral results in the IOR literature (cf \cite{tassinari93}, for similar considerations regarding keypress-measured inhibitory cueing effects). IOR is an effect that often operates on much longer time scales than those explored in the previously mentioned work supporting STD theory, and recent experimental work by Hilchey and colleagues suggests a secondary effect arising somewhere between 500 and 700 ms is responsible for the longer duration effects of IOR \cite{hilcheyUR}. Hilchey et al. performed the experiment illustrated in Fig. 2, where the time course of the effects of a transient peripheral cue were measured by way of oculomotor responses toward the cued or uncued location as commanded by either peripheral onset or central arrows signals. This work demonstrated that IOR as revealed by central arrow signals arises somewhere between 450 and 1050 ms post-cue. Furthermore, at the longest CTOA tested (1050 ms), the magnitude of the IOR effects were statistically indistinguishable when measured with peripheral and central signals, inviting the possibility that a common neural mechanism underlies the IOR effects at long CTOAs. We hypothesize that direct local inhibition of the iSC, beginning approximately 600 ms after its appearance, is responsible for this effect.

Mathematically explicit computational models are a valuable tool for generating experimentally verifiable predictions from these theories and exploring their dynamics. The dynamic neural field (DNF) model \cite{amari77,wilson73} has proven effective at modeling the dynamics of the iSC as they relate to saccade generation in a variety of paradigms \cite{kopecz95,trappenberg01,wilimzig06}. DNF models have also proven valuable in modeling IOR as understood with STD theory \cite{satel11}. Incorporating the dynamics of a DS theory of IOR with the previously implemented STD model of IOR should generate results that match those found by previous investigations using central arrow as well as peripheral targets \cite{hilcheyUR}, while also maintaining the ability to generate previous results. Here, we will use a one dimensional DNF model of the iSC to compare the simulated results of these theories of IOR. First, we will simulate the projected results of the original STD theory advocated by Satel et al. \cite{satel11} by examining the effects of peripheral cues on subsequent oculomotor responses to either peripheral onset or central arrow signals in the DNF. Second, we will test the hybrid DS theory of IOR by introducing an inhibitory signal, centered on cued locations, 600 ms after cue onset.

\begin{figure}
\centering
\includegraphics [scale=0.5]{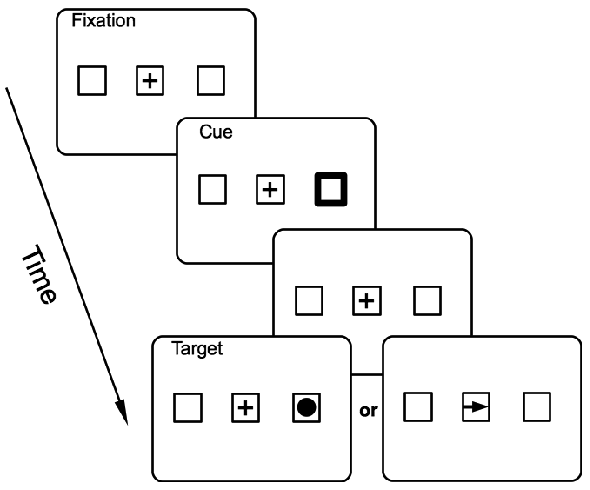}
\caption{Illustration of the simulated experimental design. The target could be either an onset dot in the peripheral or a central arrow at fixation. The CTOA was manipulated between 50 and 1050 ms.}
\end{figure}

\section{Methods}

\subsection{Experimental design}
Following Hilchey et al. \cite{hilcheyUR}, the experimental paradigm simulated is illustrated in Fig. 2. Trials begin with subjects maintaining central fixation until the appearance of the target. Spatially uninformative peripheral cues appear at various CTOAs before target onset. Targets are lateralized peripheral onset signals requiring oculomotor localization responses or left- or right-pointing central arrow stimuli commanding left- or right-ward oculomotor responses, respectively.

\subsection{Dynamic neural field model dynamics}
The iSC is responsible for the initiation of saccades to the contralateral visual field, determining direction and amplitude by the location of activity on a retinotopic motor map. A one dimensional model of the iSC was used here, where each node represents the aggregated activity of a cluster of neurons. The DNF model is characterized by short distance excitation of nearby clusters and long distance inhibition that captures the behavior of iSC neurons very well \cite{trappenberg01}. This model was initially developed to explain more general lateral interaction in neural systems \cite{amari77,wilson73}, but has been shown to be effective for modeling the dynamics of the iSC by deriving parameters from neurophysiological studies in monkeys \cite{arai99,trappenberg01}. Here, we will expand this model to further explore the STD and DS theories of IOR.

In this model, $n = 1001$ nodes were used to represent 5 mm of iSC tissue. A weighting matrix, $w_{ij} $, represents the magnitude and inhibitory or excitatory nature of the connection between two nodes, following the pattern of proximal excitation and distal inhibition according to the equation below:
\begin{equation}
w_{ij}=a*\exp(\frac{-((i-j)\Delta x)^2} {2\sigma_a^2})-b*\exp(\frac{-((i-j)\Delta x)^2} {2\sigma_b^2}) -c \, ,
\end{equation}
with $a = 72, b = 24, c = 6.4, \sigma_a = 0.6 mm$, and $\sigma_b = 1.8 mm$. At each time step the internal state of each node changes in accordance with the following relationship:
\begin{equation}
\tau \frac{du_i(t)}{dt}=-u_i(t)+ \sum_{j}w_{ij} r_j(t) \Delta x + I_i(t) + u_0 \, ,
\end{equation}
where {\it $u_0 = 0$} regulates the baseline resting activity of each node, $r_j$ is the activation level of node $j$ and is defined as follows:
\begin{equation}
r_j(t) = \frac {1}{1+exp(-\beta u_j(t) + \theta)} \,,
\end{equation}
with $\beta = 0.07$ and $\theta = 0$ used as parameters in the sigmoidal gain function. $I_k$ is the effect of the external input centered on node $i$. The iSC integrates exogenous (visual) and endogenous (goal-directed) signals and our model reflects this. Each input is represented as a Gaussian, where $d$ is the strength of the input and $\sigma_d$ is its width: 
\begin{equation}
I_k=d*exp(\frac {-((k-i) \Delta x)^2} {2\sigma_d^2}) \,.
\end{equation}

Exogenous and endogenous input signals were modeled with a width of $\sigma_{exo} = 0.7 mm$, and fixation input signals with a width of $\sigma_{endo} = 0.3 mm$. Exogenous signals used a transient strength of $d = 40$, with a decay constant of $t_{eff} = 1/\tau$. A 70 ms delay was introduced to exogenous signals to simulate early sensory processing delays. When testing the STD theory, endogenous move signals were introduced after an onset delay of 50 ms and sustained until a saccade was triggered \cite{trappenberg01}. When testing the DS theory, an inhibitory input, $I_{inh}$, was applied at the cued location with a strength of $d = 0.5$, arising 600 ms after cue onset. Target-elicited endogenous eye movement signals, $I_{endo}$, were simulated as sustained input with a 120 ms delay, with a strength varying from $d = 8$ to $d = 12$ based on the CTOA (to represent increasing temporal predictability) \cite{satel11}, and a width of $\sigma_{endo} = 0.7 mm$. SRTs were calculated as the difference between the time of target onset and the time when any node reaches 80\% of its maximal firing rate. A further 20 ms was added to account for the time taken for motor signals initiating saccades to traverse the brainstem and reach ocular muscles.

\begin{figure}
\centering
\includegraphics [scale=0.4]{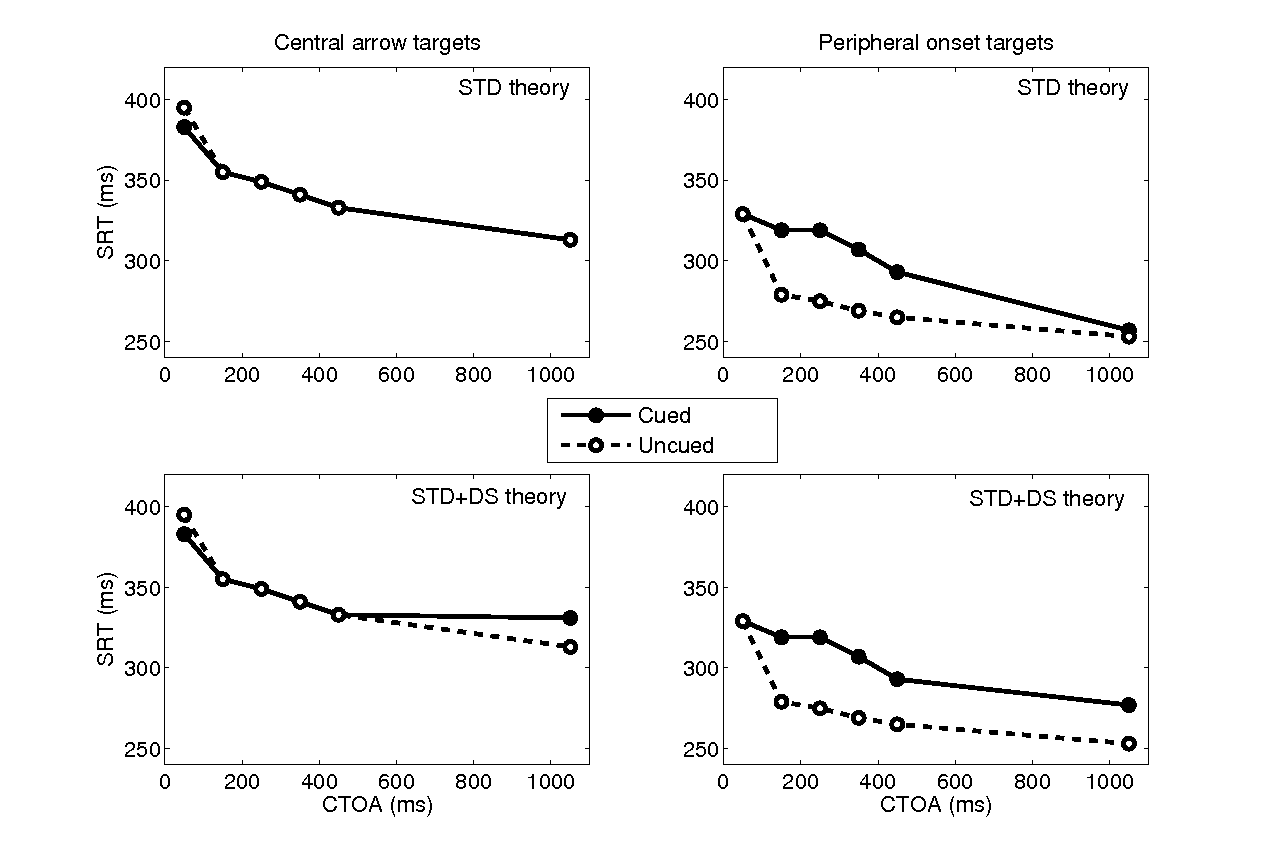}
\caption{Simulated saccadic reaction times (SRTs) for the central arrow target and peripheral arrow conditions.}
\end{figure}

\section{Results}

The STD and hybrid DS theories of IOR were tested in same (cued, valid; cue and target occur at the same location) and opposite (uncued, invalid; cue and target occur on opposite sides of the visual field) conditions with both central arrow and peripheral onset target types. The IOR scores were calculated by subtracting cued from uncued SRT for each target type condition (see Figure 4). Figure 3 illustrates the simulated SRTs for each condition, showing a general pattern that is similar to that observed in behavior.

As shown in Fig. 4A, behaviorally, IOR is only observed at late CTOAs when measured with central arrow targets, but is observed at both short and long CTOAs when peripheral onset targets are used. At short CTOAs, central targets actually led to behavioral facilitation. In the model, facilitation at short CTOAs is the result of lingering cue-elicited activation that counteracts the STD when it is generated. The STD theory alone (see Fig. 4B) can only explain IOR with repeated peripheral stimulation, and the effect decays at long intervals even though IOR is still observed. By incorporating an additional mechanism of direct inhibition that arises 600 ms after cue onset (see Fig. 1B \& 4C), the complete behavioral pattern of effects can be reproduced by the model, suggesting that STD theory alone is insufficient to explain IOR.

\begin{figure}
\centering
\includegraphics [scale=0.6]{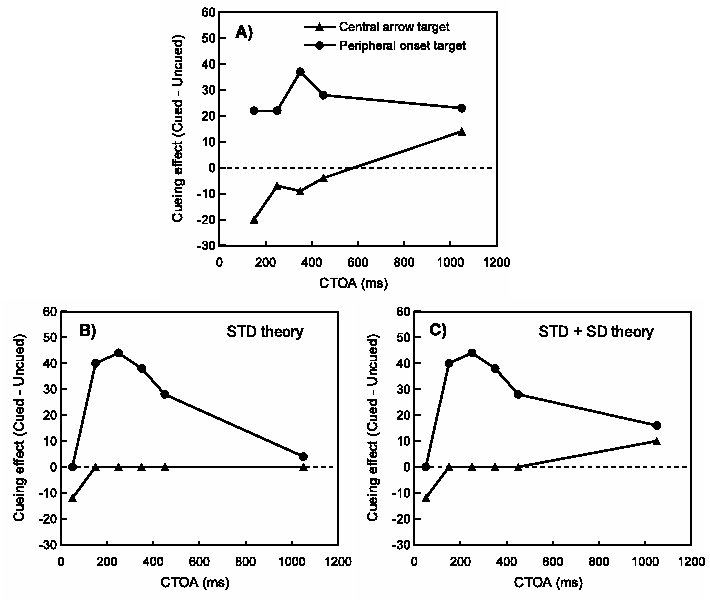}
\caption{A) Behavioral results (adapted from \cite {hilcheyUR}). B) Simulated cueing effects predicted by STD theory and C) STD+DS theory.}
\end{figure}

\section{Discussion}

The purpose of these computational models is to reach a better understanding of the dynamics of the system of interest by simulating its postulates, establishing the explanatory power of the theories with respect to currently published results, and to motivate further study by predicting behavior under unexplored conditions. Here, we have implemented a mathematically explicit model exploring the hybrid STD and DS theory of IOR. On its own, STD theory is unable to explain IOR effects without repeated stimulation, as in trials involving central arrow targets. However, the previous, unextended, STD theory of IOR \cite{satel11} is a valuable component of the updated hybrid DS model, effectively capturing cueing effects at relatively short CTOAs with repeated peripheral stimuli. The DS theory of IOR, whereby, after a processing delay, an inhibitory signal centered on the cued node reduces baseline node activity, directly increasing the stimulation required to elicit an eye movement. If it exists, as suggested by the behavioural results of Hilchey et al. \cite{hilcheyUR} (see Fig. 4A), direct collicular local inhibition must arise at some time after around 500 ms post-cue and likely modulates activity in one or more cortical regions capable of processing complex stimuli like arrows. This DS theory of IOR is capable of capturing the behavioral effects of IOR at long CTOAs, but without the incorporation of STD cannot generate an inhibitory cueing effect at short CTOAs when peripheral signals overlap in space. When combined in the hybrid STD plus DS model, simulated response times and cueing effects match the pattern of human behavioral results nicely.

\subsection{Suggestions for future research}

Further neurophysiological direct stimulation studies similar to those performed by Fecteau and Munoz \cite{fecteau05} and Dorris et. al \cite{dorris02} at CTOAs between 500 and 1000 ms in monkeys who are displaying behavioral IOR would be extremely valuable in determining the neurophysiological mechanisms underlying the hypothesized inhibitory mechanism that behavioral evidence suggests is responsible for long term IOR. If these studies revealed direct inhibition at longer CTOAs when IOR is observed, it would effectively disprove the main neurophysiological argument in favor of the STD model - namely that no direct inhibition was detected at a 200 ms CTOA when IOR was observed \cite{dorris02}. Of additional interest would be neurophysiological recordings during trials with different combinations of central arrow cues and targets, to provide better understanding of the source and nature of IOR. In a similar vein, further behavioral studies to better bound the activation latency of this secondary inhibitory input would better allow costly and time-consuming neurophysiological studies, as well as computational work, to be focused upon critical periods. Behavioral analysis at CTOAs between 500 and 1000 ms would give us a much better idea of when this signal is generated and a general examination of more time points would improve our understanding of the temporal dynamics of the mechanisms underlying IOR.

\subsection{Conclusion}

Our simulations show that the STD model is incapable of explaining IOR on its own, as it shows serious discrepancies with established behavioral data in the literature. The STD model represents early sensory effects and, in combination with facilitation, captures cueing effects quite well at short CTOAs, but cannot explain long CTOA effects, or IOR when using central arrows. When combined with a model of direct collicular local inhibition arising from cortical structures, the model is better able to reproduce established behavioral results in different paradigms. With more data from the experiments outlined above, these models could be further refined and the true neurophysiological mechanisms underlying IOR could be identified.

\section{Acknowledgments}

This work was supported by R. M. Klein's NSERC Discovery Grant. Z. Wang's participation in this project was supported by a grant from the Natural Science Foundation of Zhejiang Province, China (Grant No.: LY13C090007).

\end{document}